\documentclass[twocolumn,showpacs,preprintnumbers,amsmath,amssymb]{revtex4}

\usepackage{graphicx}
\usepackage{dcolumn}
\usepackage{bm}

\begin{document}


\title{Evidence for the Presence of Spin Accumulation in Localized States \\at Ferromagnet-Silicon Interfaces}

\author{Y. Ando,$^{1}$ S. Yamada,$^{1}$ K. Kasahara,$^{1}$ K. Masaki,$^{1}$ K. Sawano,$^{2}$ M. Miyao,$^{1}$ and K. Hamaya$^{1,3}$\footnote{E-mail: hamaya@ed.kyushu-u.ac.jp}}

\affiliation{$^{1}$Department of Electronics, Kyushu University, 744 Motooka, Fukuoka 819-0395, Japan}%
\affiliation{$^{2}$Advanced Research Laboratories, Tokyo City University, 8-15-1 Todoroki, Tokyo 158-0082, Japan}
\affiliation{$^{3}$PRESTO, Japan Science and Technology Agency, Sanbancho, Tokyo 102-0075, Japan}%

%

\date{\today}
\begin{abstract}
We experimentally show evidence for the presence of spin accumulation in localized states at ferromagnet-silicon interfaces, detected by electrical Hanle effect measurements in CoFe/$n^{+}$-Si/$n$-Si lateral devices. By controlling the measurement temperature, we can clearly observe marked changes in the spin-accumulation signals at low temperatures, at which the electron transport across the interface changes from the direct tunneling to the two-step one via the localized states. We discuss in detail the difference in the spin accumulation between in the Si channel and in the localized states. 

\end{abstract}
\maketitle


The injection of spin-polarized electrons, manipulation of the injected spins, and detection of the manipulated spins were achieved for semiconductor (SC) devices.\cite{Crooker,Crowell,Appelbaum} Thanks to these technological jumps, many advanced physical phenomena in a field of semiconductor spintronics have so far been revealed.\cite{Dery1,Huang,Crooker1,Cheng,Li} Recently, the electrical detection of spin accumulation and its depolarization in SCs through the Hanle effect in the three-terminal lateral devices has frequently been reported for GaAs,\cite{Lou,Tran} Si,\cite{Jansen,Jeon,Ando1,Ando2,Ando3,Sasaki2,Ishikawa} and Ge.\cite{Saito,Kasahara,Jain,Jeon2} Since the device geometry used in this measurement is very simple and the multiple ferromagnetic contacts or submicron-sized fabrication processes are not necessary, this method is usually utilized as an evidence for the first step of the spin injection and detection in SCs.\cite{Lou,Tran,Jansen,Jeon,Ando1,Ando2,Ando3,Sasaki2,Ishikawa,Saito,Kasahara,Jain,Jeon2} 

However, underlying physics of the spin-accumulation signals detected in the three-terminal methods has been argued. The critical issue was firstly raised by Tran {\it et al.} in Fe/Al$_\text{2}$O$_\text{3}$/GaAs structures.\cite{Tran} They reported a considerably large voltage drop ($|\Delta$$V_\text{Hanle}|$), i.e., the magnitude of spin accumulation, of $\sim$ 1 mV, which cannot be explained by the theory based on spin injection and spin diffusion.\cite{Fert,Jedema2,Takahashi,Dery2} In general, since the spin accumulation in SCs is corresponding to the difference in electrochemical potential between up and down spins, such large $|\Delta$$V_\text{Hanle}|$ cannot be created in the large density of states in the SC channels.\cite{Fert,Jedema2,Takahashi,Dery2} Eventually, they theoretically claimed that the large $|\Delta$$V_\text{Hanle}|$ originates from the spin accumulation in extrinsic localized states such as ionized donors within the depletion layer or surface states at the insulator (I)/SC interface.\cite{Tran} Very recently, Jansen {\it et al.} also explained theoretically the magnitude of spin accumulation created in SC channels by spin-polarized tunneling via interface states.\cite{Jansen2} Their study particularly showed an influence of the parallel event of direct tunneling and two-step one via the localized states on the spin accumulation, and revealed that there are some conditions that can equalize the $|\Delta$$V_\text{Hanle}|$ values derived from the SC channels and from the localized states.\cite{Jansen2} Considering these arguments, one has to discuss whether the spin accumulation occurs dominantly in the SC channels or in the localized states in the three-terminal Hanle-signal measurements. 

Using the three-terminal method in a metal-oxide-semiconductor field effect transistor (MOSFET) structure,\cite{Ando2} we recently demonstrated the electric field control of $|\Delta$$V_\text{Hanle}|$ in Si at room temperature. At least, the above features and the $|\Delta$$V_\text{Hanle}|$ values can be explained within a framework based on the simple diffusion model.\cite{Fert,Jedema2,Takahashi,Dery2} Namely, we have already observed $|\Delta$$V_\text{Hanle}|$ originating from the spin accumulation in the Si channels. In contrast, there is no clear evidence for the presence of the spin accumulation in the localized states, where the study by Tran {\it et al.} is not experimental evidence but is a possible comment so as to explain their data.\cite{Tran} 

In this Letter, we experimentally show the direct evidence for the presence of spin accumulation in the localized states at ferromagnet-silicon interfaces. By controlling the measurement temperature, we can clearly observe marked changes in the Hanle-effect signals at low temperatures, at which the transport mechanism of the electrons changes from the direct tunneling to the two-step one via the localized states. Since the Schottky-tunnel contact of our devices has atomically controlled heterointerfaces\cite{Hamaya1,Maeda} and we do not use insulating tunnel barriers, the influence of surface states at the I/SC interface should be excluded. Thus, we can infer that the localized states arise from the ionized Sb, used for modulation doping in the contact region, within the depletion layer. We also discuss the difference in the Hanle-effect signals derived from the spin accumulation between in the Si channel and in the localized states. 
\begin{figure}[t]
\includegraphics[width=7.5cm]{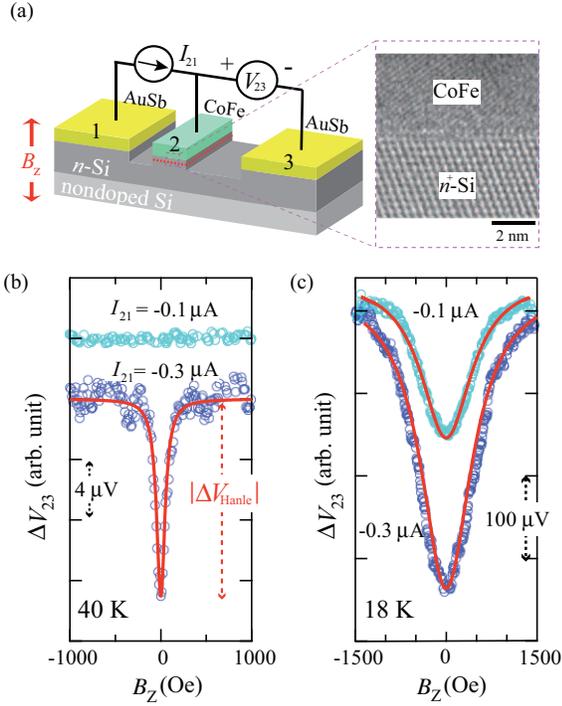}
\caption{(Color online) (a) Schematic diagram of a Si-based three-terminal device with a CoFe/$n^{+}$-Si Schottky-tunnel contact fabricated. The enlarged picture is a cross-sectional transmission electron microscopy image of the CoFe/$n^{+}$-Si/$n$-Si heterostructure. (b) and (c) are $\Delta$$V_\text{23}$-$B_\text{Z}$ curves at 40 K and 18 K, respectively. The solid curves are fitting results with Lorentzian function.\cite{Jansen} }
\end{figure}

The three-terminal devices with a CoFe/$n^{+}$-Si Schottky-tunnel contact were fabricated as follows. First, $\sim$ 200-nm-thick $n$-Si channel with a carrier density ($n$) of $ \sim$ 6.0 $\times$ 10$^{17}$ cm$^{-3}$ at  room temperature was formed on an undoped FZ-Si(111) substrate by an ion implantation (phosphorus) technique. Next, $n^{+}$-Si layer with a thickness of $\sim$ 7 nm was grown on top of the $n$-Si channel layer by a combination of the Si epitaxy using an MBE process with an Sb $\delta$-doping technique (Sb : 1 $\times$ 10$^{19}$cm$^{-3}$).\cite{Nakagawa,Miyao,Sawano,Hamaya1} Because of  the surface segregation of doped Sb atoms in the Si epilayer, the doping profile of Sb can be modulated in the $n^{+}$-Si layer.\cite{Nakagawa,Miyao} Although the carrier concentration at the Sb $\delta$-doped position is $n$ $\sim$ 10$^{19}$cm$^{-3}$, that near the surface becomes $n$ $\sim$ 10$^{18}$cm$^{-3}$. After the chemical etching of the surface ($\sim$ 2 nm), we grew 10-nm-thick CoFe epitaxial layer on top of it by low-temperature molecular beam epitaxy (LT-MBE) at 60 $^{\circ}$C.\cite{Maeda} As a result, a Schottky-tunnel contact consisting of CoFe(10 nm)/$n^{+}$-Si($\sim$ 5 nm)/$n$-Si was realized, where the $n^{+}$-Si contact region has a modulated doping profile, i.e., $\sim$ 10$^{18}$cm$^{-3}$ $\to$ $\sim$ 10$^{19}$cm$^{-3}$ from the top. By using conventional processes with photolithography, Ar$^{+}$ ion milling, and reactive ion etching, the three-terminal lateral device structure is defined. Also, two ohmic contacts (AuSb) were formed at less than 250 $^{\circ}$C. Here the $n$$^{+}$-Si layer on the channel region was removed by the Ar$^{+}$ ion milling. A schematic illustration of the fabricated devices is shown in Fig. 1(a). The contact 2 has a lateral dimension of 6 $\times$ 200 $\mu$m$^{2}$ and the distance between the contacts 2 and 1 or 3 is $\sim$ 50 $\mu$m or $\sim$ 70 $\mu$m, respectively. As displayed in the enlarged TEM picture, the fabricated CoFe/$n^{+}$-Si/$n$-Si structure has an atomically flat interface. Thus, we can ignore defects and interface states arising from the interface reaction. The three-terminal Hanle-effect measurements were performed by a dc method with the current ($I_\text{21}$) $-$ voltage ($V_\text{23}$) configuration shown in Fig. 1(a). In the measurements, a small magnetic field perpendicular to the plane, $B_\text{Z}$, was applied after the magnetic moment of the contact 2 aligned parallel to the plane along the long axis of the contact.
\begin{figure}[t]
\includegraphics[width=7.5cm]{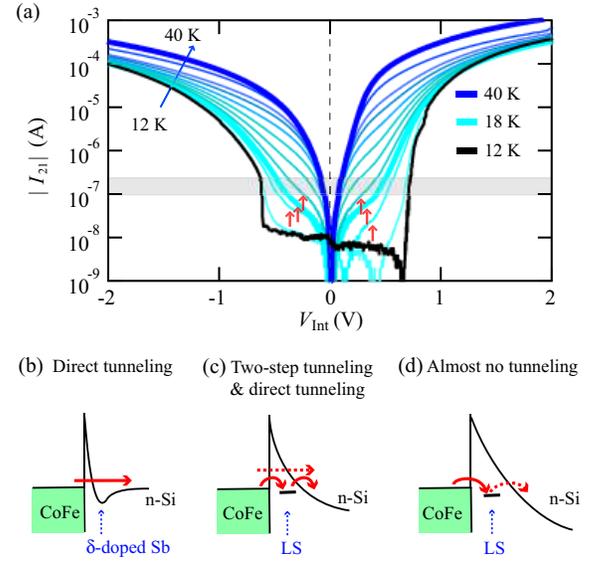}
\caption{(Color online) (a) $|I_\text{21}|$$-V_\text{Int}$ characteristics from 12 to 40 K. (b), (c), and (d) are schematic illustrations of electron transport via direct tunneling, two-step tunneling with weak direct tunneling, and almost no tunneling, respectively, in reverse bias conditions ($V_\text{Int} < 0$). To obtain the same tunnel current, we have to increase the $V_\text{Int}$ value with decreasing temperature. Thus, each schematic illustration indicates different $V_\text{Int}$ conditions. }
\end{figure} 

We first measured the three-terminal voltage changes ($\Delta$$V$$_\text{23}$) as a function of $B_\text{Z}$, i.e., $\Delta$$V_\text{23}$-$B_\text{Z}$ curves, where quadratic background voltages depending on $B_\text{Z}$ are subtracted from the raw data. Here the electrons are injected into and extracted from, respectively, the Si conduction band for reverse ($I <$ 0) and forward ($I >$ 0) biases, as previously shown.\cite{Ando1,Ando2,Ando3} Figures 1(b) and 1(c) show $\Delta$$V_\text{23}$-$B_\text{Z}$ curves for $I_\text{21}$ $=$ -0.1 and -0.3 $\mu$A at 40 and 18 K, respectively. At 40 K [Fig. 1(b)], we can clearly see the bias-dependent spin detection, as previously discussed in Ref.\cite{Ando1}. That is, although the spin signal can not be detected in a low reverse bias condition ($I_\text{21}$ $=$ -0.1 $\mu$A), the clear spin signal, $|\Delta$$V_\text{Hanle}|$$\sim$ 13 $\mu$V, appears suddenly by an increase in $|I$$_\text{21}|$ ($I_\text{21}$ $=$ -0.3 $\mu$A). This feature means that the available state of the Si conduction band near the quasi Fermi level can become spin-polarized by the spin injection from the CoFe contact, consistent with our previous works on the spin accumulation in the Si channels.\cite{Ando1,Ando2,Ando3} On the other hand, at 18 K [Fig. 1(c)], we can find marked change in features of $\Delta$$V_\text{23}$-$B_\text{Z}$ curves despite the same device. In particular, $|\Delta$$V_\text{Hanle}|$ is significantly enhanced to $\sim$ 350 $\mu$V for $I_\text{21}$ $=$ -0.3 $\mu$A. Also, the large spin signal can be observed even for $I_\text{21}$ $=$ -0.1 $\mu$A where the spin signal could not be detected at 40 K. It seems that the full width at half maximum of the $\Delta$$V_\text{23}$-$B_\text{Z}$ curve increases. Using a simple Lorentzian function, $\Delta$$V_\text{23}$($B_\text{Z}$) $=$ $\Delta$$V_\text{23}(0)$/[1+($\omega_\text{L}$$\tau_\text{S}$)$^{2}$],\cite{Jansen} we can evaluate the lower limit of spin lifetime ($\tau_\text{S}$) from the $\Delta$$V_\text{23}$-$B_\text{Z}$ curves.  Here $\omega_\text{L} =$ $g\mu_\text{B}$$B_\text{Z}$/$\hbar$ is the Larmor frequency, $g$ is the electron $g$-factor ($g =$ 2), $\mu_\text{B}$ is the Bohr magneton. The fitting results are denoted by the solid curves in Figs. 1(b) and 1(c). For $I_\text{21}$ $=$ -0.3 $\mu$A, we obtained a large difference in $\tau_\text{S}$ of 1.10 nsec and 0.125 nsec at 40 and 18 K, respectively. One has never seen these marked differences in the Hanle signals in the same device.

To understand the above strange features depending on temperature, we examine the detailed $I - V$ characteristics of the fabricated CoFe/$n^{+}$-Si contact for various temperatures. Figure 2(a) shows $|I_\text{21}|$$-V_\text{Int}$ curves from 40 to 12 K every 2 K, where $V_\text{Int}$ is the bias voltage at the CoFe/Si interface and $V_\text{Int} < 0$ represents the spin injection condition. Surprisingly, $|I_\text{21}|$$-V_\text{Int}$ feature clearly changes with deceasing temperature. For all the temperatures, there is almost no rectification in large bias regime ($|V_\text{Int}|$ $>$ 1.5 V), indicating that tunneling conduction of electrons through the Schottky-tunnel barrier is dominant. However, we can find marked variation in the current in low bias regime (-0.6 V $<$ $V_\text{Int}$ $<$ 0.6 V). It should be noted that in the low bias regime two-step features can be clearly seen at around 18 K (see-arrows). These features mean that the dominant transport mechanism of electrons is changed from direct tunneling to two-step one by decreasing temperature in both bias conditions. 
\begin{figure}[t]
\includegraphics[width=7.5cm]{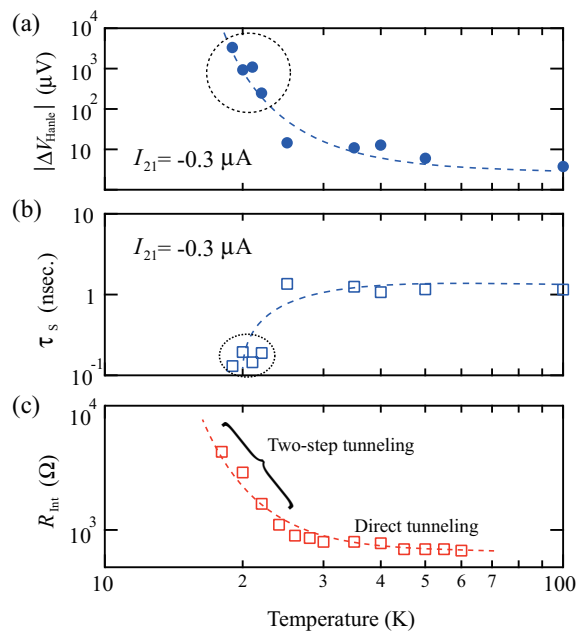}
\caption{(Color online) Temperature dependence of (a) $|\Delta$$V_\text{Hanle}|$, (b) $\tau_\text{S}$ and (c) interface resistance ($R_\text{Int}$). }
\end{figure} 

In general, the two-step tunneling of electrons occurs via interface states such as ionized donors within the depletion layer or surface states at the I/SC interface.\cite{Tran,Jansen2} Since we do not use insulating tunnel barriers between CoFe and Si and the used Schottky-tunnel contact of our devices has atomically controlled heterointerfaces,\cite{Hamaya1,Maeda} the two-step tunneling via the surface localized states at the interface should be excluded. Thus, we should take into account the localized states (LS) induced by ionized donors within the depletion layer. Considering the presence of the heavily doped Sb atoms between CoFe and Si, we can schematically illustrate the variation in the electron transport with decreasing temperature, as shown in Figs. 2(b)-(d). As described in the experimental section, the $n^{+}$-Si contact region of our three-terminal device has a modulated doping profile, $\sim$ 10$^{18}$cm$^{-3}$ $\to$ $\sim$ 10$^{19}$cm$^{-3}$, from the top. At around 40 K, the electron transport is governed by the direct tunneling across the thin tunnel barrier due to the presence of the $n^{+}$-Si layer [Fig. 2(b)]. Since the carrier density in our $n$-Si channel at 40 K was $\sim$ 2.0 $\times$ 10$^{15}$ cm$^{-3}$, determined by Hall-effect measurements, the carrier freeze-out in the $n$-Si channel can occur with further decreasing temperature.\cite{Pearson} Simultaneously, the top region in the Sb-doped Si layer can also show the carrier freeze-out because the carrier density in this region is almost equal to the channel one. As a consequence of the expansion of the width of the depletion layer with decreasing temperature, the direct tunneling tends to be suppressed and the two-step tunneling via the localized states becomes dominant, as shown in Fig. 2(c). At this time, the localized states arise from the ionized Sb atoms with a doping concentration of $\sim$ 10$^{19}$cm$^{-3}$. At around 12 K, even the two-step tunneling is almost suppressed, giving rise to the suppression of the tunnel current in the low bias regime, as illustrated in Fig. 2(d) (see-the plateau in the $|I_\text{21}|$$-V_\text{Int}$ curve at 12 K). When the large bias voltage $V_\text{Int}$ is applied ($\sim$ 0.6 V), Fowler-Nordheim tunneling takes place,\cite{FN} leading to the sudden enhancement in the current.

Considering Figs. 2(b)-(d), we discuss the correlation among $|\Delta$$V_\text{Hanle}|$, $\tau_\text{S}$, and tunneling mechanism of spin-polarized electrons. Figures 3(a)-(c) show results of temperature dependence of $|\Delta$$V_\text{Hanle}|$, $\tau_\text{S}$ and interface resistance ($R_\text{Int}$), respectively, where the $R_\text{Int}$ variation means the change in the dominant mechanism of electron transport. As shown in Fig. 3(c), the enhancement in $R_\text{Int}$ begins at around 20 K, indicating that the two-step tunneling governs the electron transport across the Schottky tunnel barrier. Simultaneously, the $|\Delta$$V_\text{Hanle}|$ values at around 20 K are markedly enhanced up to two or three orders of magnitude larger than those at more than 25 K [see-Fig. 3(a)]. Taking the change in $R_\text{Int}$ with temperature variation into account, we can regard the marked change in $|\Delta$$V_\text{Hanle}|$ as a consequence of the change in the transport mechanism of spin-polarized electrons. That is, when the spin-polarized electrons are injected predominantly into the Si channel by the two-step tunneling, the spin signals induced by the spin accumulation are markedly enhanced. As a result, this is   the direct observation of the correlation between the spin-accumulation signals in the three-terminal method and the tunneling mechanism of spin-polarized electrons at FM-Si interfaces. 
\begin{figure}[t]
\includegraphics[width=7.5cm]{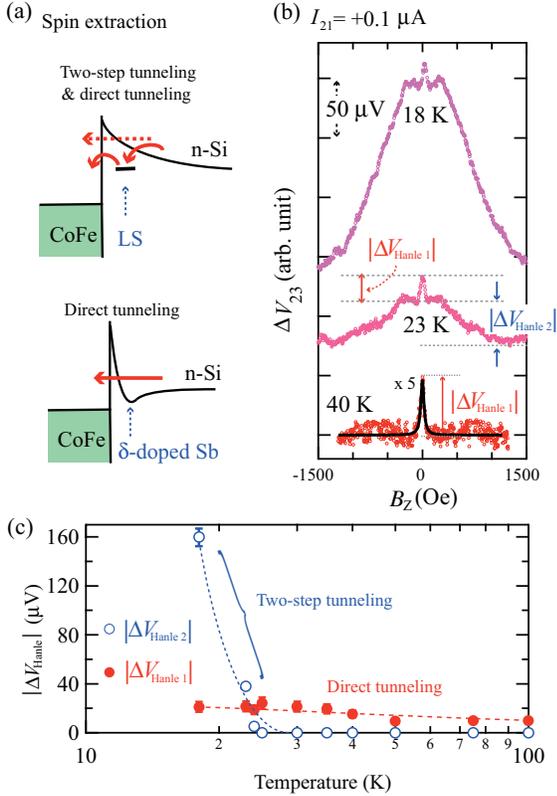}
\caption{(Color online) (a) $\Delta$$V_\text{23}$-$B_\text{Z}$ curves for $I_\text{21}$ $=$ + 0.1 A at 40, 23, and 18 K. The red solid curve at 40 K is fitting result with Lorentzian function.\cite{Jansen} (b) Schematic illustrations of electron transport via two-step tunneling with weak direct tunneling (top) and direct tunneling (bottom) in forward bias conditions ($V_\text{Int} > 0$). (c) $|\Delta$$V_\text{Hanle 1}|$ and $|\Delta$$V_\text{Hanle 2}|$ as a function of temperature. }
\end{figure}

Here, we focus on $\tau_\text{S}$. Comparing Figs. 3(b) and 3(c), we can find that $\tau_\text{S}$ is significantly suppressed by the appearance of the two-step tunneling. Actually, $\tau_\text{S}$ for the direct tunneling is $\sim$ 1.0 nsec while that for the two-step tunneling is $\sim$ 0.1 nsec. It is quite strange that the markedly large spin signals show the shorter spin relaxation time in the same device. Thus, we should further discuss the results in Figs. 3. Recent theoretical study by Jansen {\it et al.} suggested that it is important to understand a weighted average of the spin accumulation in the channel and that in the localized states.\cite{Jansen2} Following their considerations, we can speculate that the observed features at around 20 K is almost different characteristics from the features in the channels. As Tran {\it et al.} previously discussed,\cite{Tran} if we can also regard the quite large $|\Delta$$V_\text{Hanle}|$ as the spin-accumulation signals dominantly from the localized states, the above strange feature can tentatively be explained. As discussed in Fig. 2, the localized states in our device consist mainly of the heavily doped Sb atoms.\cite{Ando1,Ando2,Ando3,Hamaya1} In general, Sb has a large atomic number and has a larger spin-orbit interaction than the P-doped Si channel used. If the spin polarized electrons accumulate in the localized states formed from Sb atoms, one should ignore the weak spin-orbit interactions from the Si matrix to the injected spins. That is, the injected spins in the localized states can directly interact with the doped Sb atoms. As a result, the very fast depolarizations can occur in the device we fabricated at around 20 K. We conclude that the correlation between the enhancement in $|\Delta$$V_\text{Hanle}|$ and the suppression of $\tau_\text{S}$ can be understood by taking into account the spin accumulation in the localized states.

We finally present the three-terminal Hanle-effect signals in the spin extraction conditions ($I >$ 0). As expected in Fig. 2(a), we can detect the two-step tunneling via the localized states even in $V_\text{Int} >$ 0, schematically shown in Fig. 4(a). The direct tunneling can work as the spin extraction from the Si channel at around 40 K, while the two-step tunneling can also create the spin accumulation in the localized states at around 20 K. Figure 4(b) shows $\Delta$$V_\text{23}$-$B_\text{Z}$ curves in a spin extraction condition of $I_\text{21}$ $=$ +0.1 $\mu$A at various temperatures. Very interestingly, we can observe a double Hanle-like shape which includes a normal Hanle-effect signal in the low-field region at less than 23 K, although a normal Hanle signal can be detected at 40 K. Since the general magnitude of $|\Delta$$V_\text{Hanle}|$ ($\sim$ 10 $\mu$V) and  $\tau_\text{S}$ value ($\sim$ 1.0 nsec) are observed, as reported in our previous works,\cite{Ando1,Ando2,Ando3} the data detected at 40 K arises from the spin accumulation in the Si channel. On the other hand, the double Hanle-like shape can appear at around 20 K, at which the electron transport across the interface changes from the direct tunneling to the two-step one. Hence, these features are also related to the presence of the spin accumulation in the localized states. Since we have shown the significant electrical detectability of the spin accumulation in Si channels in the spin extraction conditions with a small bias current,\cite{Ando1,Ando3} the observed double Hanle-like shape is a peculiar feature for our device with a clear bias-current dependence. 

If we tentatively try to separate the three-terminal signals into $|\Delta$$V_\text{Hanle 1}|$ and $|\Delta$$V_\text{Hanle 2}|$, as shown in Fig. 4(b) (see- the data at 23 K), we can observe the temperature dependence of both signals in Fig. 4(c). As a result, whereas $|\Delta$$V_\text{Hanle 2}|$ appears at less than 23 K, $|\Delta$$V_\text{Hanle 1}|$ is gradually increased even at less than 23 K. This feature means that the spin accumulation in the Si channel can also be enhanced even if there is a two-step tunneling of electrons via the localized states. The temperature evolution of the two distinct spin accumulations indicates that the spin accumulations in the Si channel and in the localized states are parallel event.\cite{Jansen2} Since the $n$-Si channels for our devices have relatively low carrier density $n <$ $\sim$ 10$^{15}$ cm$^{-3}$ at low temperatures, we can create relatively large spin accumulation in the channel by using a small extraction current.\cite{Ando1,Ando3} However, if we use a heavily doped channel ($n >$ $\sim$ 10$^{19}$ cm$^{-3}$ at low temperatures), we cannot create such large spin accumulation ($|\Delta$$V_\text{Hanle}|$ $\sim$ 10 $\mu$V) comparable to the spin accumulation in the localized states ($|\Delta$$V_\text{Hanle}|$ $\sim$ 100 $\mu$V) by using the three-terminal method. Therefore, the observed peculiar features in this study are original characters of our devices with two-step tunneling via the localized states in our devices.  

In summary, we experimentally showed the evidence for the presence of spin accumulation in the localized states at ferromagnet-silicon interfaces, detected by electrical Hanle effect measurements in CoFe/$n^{+}$-Si/$n$-Si three-terminal lateral devices. The enhancement in the spin-accumulation signals and the suppression of the spin lifetime can be observed simultaneously. The observed features are original characters of our Si channel with relatively low carrier density, leading to the relatively large spin accumulation ($\sim$ 10 $\mu$V) comparable to the spin accumulation in the localized states ($\sim$ 100 $\mu$V) in the three-terminal method.

\vspace{5mm}
This work was partly supported by PRESTO-JST and STARC. Three of the authors (Y.A. K.K. and S.Y.) acknowledge JSPS Research Fellowships for Young Scientists. 


\end{document}